\documentclass[twocolumn]{aastex631}

\usepackage{CJKutf8}

\begin{document}

\title{An extremely low-density exoplanet spins slow}

\correspondingauthor{Wei Zhu}
\email{weizhu@tsinghua.edu.cn}

\author[0009-0007-6412-0545]{Quanyi Liu \begin{CJK*}{UTF8}{gbsn}(刘权毅)\end{CJK*}}
\affiliation{Department of Astronomy, Tsinghua University, Beijing 100084, China}

\author[0000-0003-4027-4711]{Wei Zhu \begin{CJK*}{UTF8}{gbsn}(祝伟)\end{CJK*}}
\affiliation{Department of Astronomy, Tsinghua University, Beijing 100084, China}

\author[0000-0003-1298-9699]{Kento Masuda \begin{CJK*}{UTF8}{min}(増田賢人)\end{CJK*}}
\affiliation{Department of Earth and Space Science, Graduate School of Science, Osaka University, 1-1 machikaneyama, Toyonaka, Osaka 560-0043, Japan}

\author[0000-0002-2990-7613]{Jessica E. Libby-Roberts}
\affiliation{Department of Astronomy \& Astrophysics, 525 Davey Laboratory, The Pennsylvania State University, University Park, PA 16802, USA}
\affiliation{Center for Exoplanets and Habitable Worlds, 525 Davey Laboratory, The Pennsylvania State University, University Park, PA 16802, USA}

\author[0000-0003-3355-1223]{Aaron Bello-Arufe}
\affiliation{Jet Propulsion Laboratory, California Institute of Technology, Pasadena, CA 91011, USA}

\author[0000-0003-4835-0619]{Caleb I. Cañas}
\affiliation{NASA Goddard Space Flight Center, 8800 Greenbelt Road, Greenbelt, MD 20771, USA}

\begin{abstract}
We present constraints on the shape of Kepler-51d, which is a super-puff with a mass $\sim6\,M_\oplus$ and a radius $\sim9\,R_\oplus$, based on detailed modeling of the transit light curve from JWST NIRSpec. 
The projected shape of this extremely low-density planet is consistent with being spherical, and a projected oblateness $f_\perp>0.2$ can be excluded regardless of the spin obliquity angles. If this is taken as the limit on the true shape of the planet, Kepler-51d is rotating at $\lesssim 50\%$ of its break-up spin rate, or its rotation period is $\gtrsim 33\,$hr. In the more plausible situation that the planetary spin is aligned with its orbital direction to within $30^\circ$, then its oblateness is $<0.08$, which corresponds to a dimensionless spin rate $\lesssim30\%$ of the break-up rotation and a dimensional rotation period $\gtrsim 53\,$hr. This seems to contradict the theoretical expectation that planets with such low masses may be spinning near break-up. 
We point out the usefulness of the stellar mean density and the orbital eccentricity in constraining the shape of the transiting planet, so planets with well-characterized host and orbital parameters are preferred in the detection of planetary oblateness with the JWST transit method.

\end{abstract}

\section{Introduction} \label{sec:intro}

The spin state of a planet conveys useful information on the formation and evolution of the planet, as has been shown by studies of the Solar System planets \citep[e.g.,][]{Ward:2004, Hamilton:2004, Ida:2020}. As the mass growth is primarily through the accretion from the surrounding material that carries angular momentum, the newly formed planets are expected to spin at a significant fraction or even close to the break-up spin rate, $\Omega_{\rm brk} \equiv \sqrt{GM_{\rm p}/R_{\rm p}^3}$, irrespective of the planet mass \citep[e.g.,][]{Zhu:2015, Batygin:2018, Dong:2021, Takaoka:2023}. Here $M_{\rm p}$ and $R_{\rm p}$ are the planetary mass and (mean) radius, respectively. After the formation, dynamical evolution at later stage can also alter the spin state substantially \citep[e.g.,][]{Ward:2004, Hamilton:2004, LiLai:2020}. 

Fast-spinning planets are flattened, and the oblateness/flattening parameter is commonly used to measure the shape distortion, which is defined as
\begin{equation}
    f \equiv \frac{R_{\rm eq} - R_{\rm pol}}{R_{\rm eq}} .
\end{equation}
Here $R_{\rm eq}$ and $R_{\rm pol}$ are the equatorial and polar radii, respectively. This oblateness parameter can then be related to the spin rate $\Omega$ via the Darwin--Radau relation \citep[e.g.,][Chapter 4]{Murray:1999}
\begin{equation} \label{eqn:darwin}
    \left( \frac{\Omega}{\Omega_{\rm brk}} \right)^2 = \left[ \frac{5}{2}\left(1-\frac{3}{2}\bar{C}\right)^2 + \frac{2}{5} \right] f ,
\end{equation}
where $\bar{C}$ is the moment of inertia of the planet in unit of $M_{\rm p}R_{\rm p}^2$. A spherical planet with uniform density has $\bar{C}=0.4$, whereas for Solar System giants, which are more concentrated in the core, their moments of inertia are in the range of 0.2--0.3 \citep[e.g.,][]{Ni:2018, Jack:2022, Neuenschwander:2022}.

The oblate shape of a planet can in principle be detected through high-precision transit light curve \citep{Seager:2002, Hui:2002, Barnes:2003}. The oblateness-induced signal, defined as the difference in the transit light curve from an oblate planet and a spherical planet with the same cross-sectional area, appears almost entirely during the ingress and the egress and has an amplitude $\lesssim 200\,$ppm for a Jupiter-sized ($R_{\rm p}/R_\star\approx0.1$) planet with a Saturn-like oblateness ($f\approx0.1$, \citealt{Carter:2010, Zhu:2014}). Together with the relatively short ($\lesssim$hr) ingress and egress durations, constraining the shape of a transiting planet has been very challenging.

The James Webb Space Telescope (JWST) has demonstrated excellent photometric performance in transit observations, achieving $\lesssim$100\,ppm precision with only a few minutes exposures on relatively bright targets \citep[e.g.,][]{Ahrer:2023, Alderson:2023, Rustamkulov:2023, Feinstein:2023}. In addition, the near-infrared coverage of JWST is able to further boost the signal-to-noise ratio with reduced limb-darkening effect and stellar activities. As has been demonstrated in \citet{Liu:2024}, the oblateness signals of over one dozen known transiting planets with relatively long orbital periods and deep transits can already be detected, or at least constrained to a physically meaningful level, with JWST transit observations.

In this paper, we report the constraints on the oblateness of Kepler-51d from modeling the JWST transit light curve. Kepler-51d orbits a young G-type star of age $\sim$500\,Myr and $J$ magnitude 13.56 with an orbital period of 130 days \citep{Libby-Roberts:2020}. With a measured density of $\sim0.05\,{\rm g\,cm^{-3}}$ \citep{Masuda:2014}, Kepler-51d joins the rare class of extremely low-density ($<0.1\,{\rm g\,cm^{-3}}$) planets called super-puffs, whose formation and evolution remain to be understood \citep{Lee:2016, Millholland:2019, WangDai:2019}. For Kepler-51d in particular, this planet resides in a system with three super-puffs in a near-resonance configuration, and its transmission observations by Hubble Space Telescope (HST) have yielded featureless spectra \citep{Libby-Roberts:2020}.

Kepler-51d is a particularly interesting target for planetary oblateness studies. First, its deep transit and long orbital period allow us to verify with real data the detectability of planetary oblateness with JWST. With an orbital period of $\sim 130$ d, this planet is not expected to be tidally spin-down within the relatively young age of the system \citep[e.g.,][]{Murray:1999}, so the oblate shape is related to the planetary spin and thus we can constrain the spin rate based on the oblateness measurement/constraint. Compared to gas giants, low-mass planets that do not open up gaps in the protoplanetary disk have different formation pathways, which may result in different spin states. Normally it would be difficult, if not impossible, to constrain the shapes and spins of low-mass planets even with JWST, but super-puffs like Kepler-51d allow us to do this. Furthermore, the extremely low density of Kepler-51d means that the planet can be more easily flattened by rotation. Specifically, its break-up spin period is about 16 hr, which is similar to the actual spin period of Uranus or Neptune and much longer than their break-up spin period.

This paper is organized as follows. Section~\ref{sec:data} summarizes the JWST observations and the data reduction method, Section~\ref{sec:model} describes the modeling procedures, Section~\ref{sec:result} presents the results of our oblateness modeling, and Section~\ref{sec:discussion} discusses the implications of our results and the difference between our work and the independent work by \citet{Lammers:2024}, which also studied the oblateness and spin of Kepler-51d based on the same JWST dataset. Finally in Section~\ref{sec:summary} we briefly summarize our findings.

\section{Observations \& Data Reduction} \label{sec:data}

\begin{figure*}
    \centering
    \includegraphics[width=\linewidth]{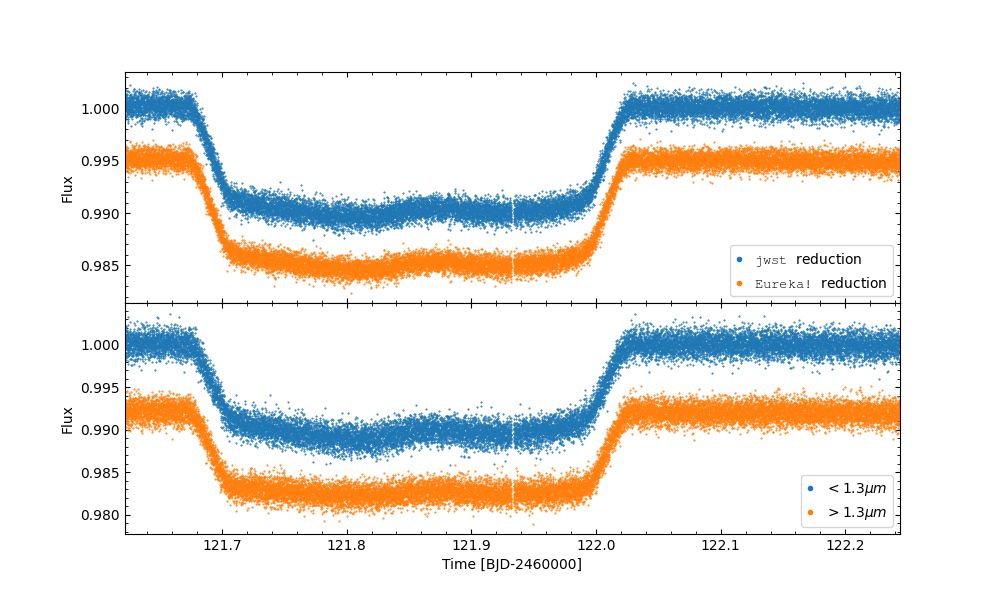}
    \caption{Light curves from different reductions (upper panel) and different wavelength channels (lower channel). The out-of-transit variations of \texttt{jwst} and \texttt{Eureka!} reduction are $\sim 650$ and $\sim 580\,$ppm respectively, and $<1.3\mu$m and $>1.3\mu$m channels are $\sim 950$ and $\sim 880\,$ ppm. A spot-crossing event occurred near the mid-transit. Its amplitude decreases with increasing wavelength, as expected.}
    \label{fig:all_lc}
\end{figure*}

JWST observations of a single Kepler-51d transit were obtained on June 26, 2023, through the Cycle 1 program with number GO-2571 (PI: J.~Libby-Roberts). These observations were conducted under the Bright Object Time Series (BOTS) mode with the NIRSpec instrument \citep{Birkmann:2022}. Given the faintness of the target star, CLEAR filter and PRISM disperser were chosen, and the SUB512 subarray with the NRSRAPID readout pattern were used. The science data had a duration of $\sim$15 hours, which consisted of 18,082 integrations, each with 12 groups and 2.9\,s exposure time. 

We reduce the JWST data based on the JWST calibration pipeline, \texttt{jwst} (version 1.15.1, \citealt{jwst_pipeline}), following closely the no.\ 29 JWebbinars.
\footnote{\url{https://www.stsci.edu/jwst/science-execution/jwebbinars}}
Starting from the uncalibrated files, we used the default parameters to perform the detector-level corrections and determine the flux per integration (i.e., Stage 1), except for the removal of the cosmic ray effect, for which we have adopted a higher threshold (8$\sigma$) for the ramp-jump detection. The $1/f$ noise is mitigated at the group level (see also \citealt{nirspec_ers}).
At Stage 2, the flat fielding and photometric calibration are skipped, as they are irrelevant for TSO observations. A custom 2D spectral cutout step was performed, and the bad pixels and outliers were identified and replaced following the method of \citet{Nikolov:2014}.

We use 1D Gaussians and quadratic polynomials in the spectral tracing. An aperture radius of three pixels is used to extract the 1D spectra. The extracted 1D spectra are then combined into a white light curve by summing the flux across the full wavelength range (0.55--5.35\,$\mu$m). Our reduction yields a white light curve with an out-of-transit variation of $\sim 650\,$ppm. 
In addition to the white light curve, we have also created broadband light curves in two channels with wavelength ranges 0.55--1.30\,$\mu$m and 1.30--5.35\,$\mu$m, respectively, which roughly equally divide the total stellar flux. These two-channel light curves will be used later in the consistency check of the oblateness constraints across wavelengths (see Section~\ref{sec:result}). The two-channel light curves and the combined white light curve are shown in Figure~\ref{fig:all_lc}.

The white light curve that was reduced by the \texttt{Eureka!} package \citep{Bell:2022} and used in the analysis of Libby-Roberts et al. (2024, in-prep) and \citet{Masuda:2024} is also used here as an independent check on the oblateness constraints. We refer to Libby-Roberts et al. (2024, in prep) for details about the data reduction steps. This white light curve, also shown in Figure~\ref{fig:all_lc}, has an out-of-transit variation of 580\,ppm per integration.

The JWST transit of Kepler-51d shows a $\sim2\,$hr offset in mid-transit time compared to the prediction based on the 3-planet model \citep{Libby-Roberts:2020}. This has motivated the search for the fourth planet in the same system, which yielded updated physical and orbital properties of the transiting planets, including Kepler-51d \citep{Masuda:2024}. These updated parameters are used in the present work.

%%%%%%%%%%%%%%%%%%%%%%%%%%%
\section{Oblateness Modeling} \label{sec:model}

\begin{figure*}
    \centering
    \includegraphics[width=\linewidth]{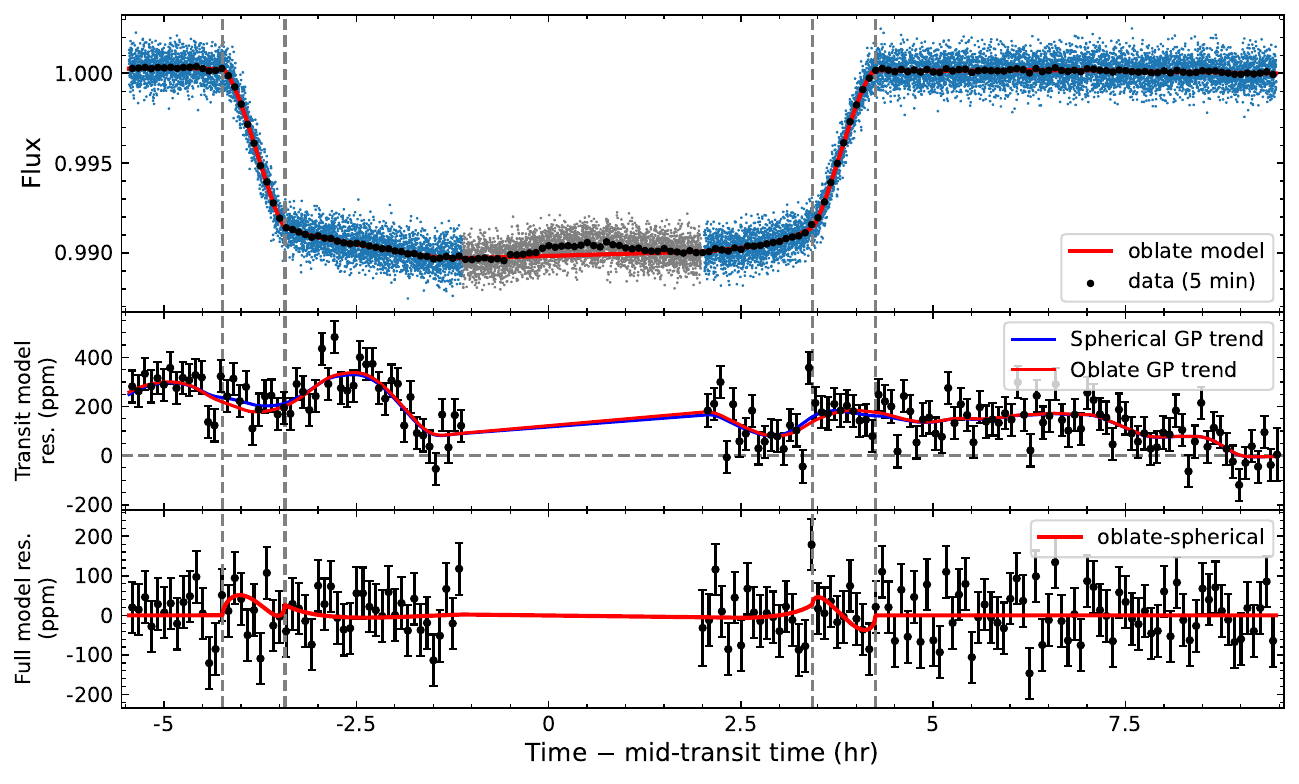}
    \caption{The white light curve and the best-fit oblate model of Kepler-51d. The top panel shows the original data points (blue and gray), the 5\,min-binned data points (black), and the best-fit oblate model (red). Data points affected by the spot-crossing event are shown in gray and excluded in the modeling. The middle panel shows the residuals after subtracting the best-fit spherical model, thus highlighting the systematic trend in the original light curve. The bottom panel shows the residuals after subtracting the full model (including both spherical model and the systematic trend), to highlight the detectability of the oblateness signal. The vertical dashed lines in all panels indicate the four contact points according to the best-fit oblate model. The oblate model used here has $f_\perp=0.05$ and $\theta_\perp=55^\circ$.}
    \label{fig:lightcurve_baseline}
\end{figure*}

\begin{figure*}
    \centering
    \includegraphics[width=\textwidth]{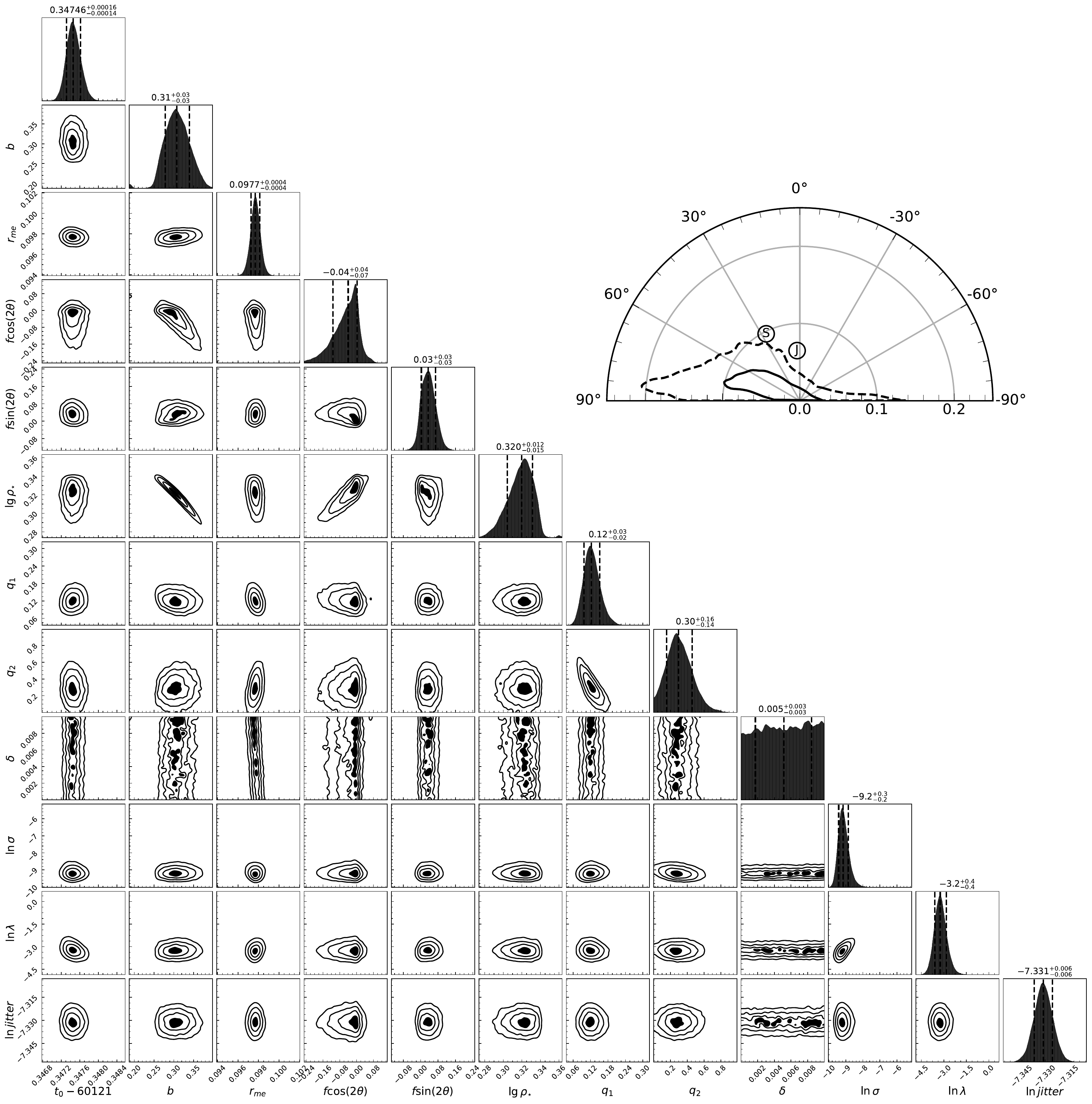}
    \caption{The corner plot of the baseline model. The values and 1-$\sigma$ uncertainties of the modeling parameters are indicated at the top of the individual column. These are defined as the median and the values enclosing $68\%$ credible interval of the marginal probability distribution, respectively. The reconstructed posterior distribution in the $f_\perp$ vs.\ $\theta_\perp$ plane is shown in the upper right corner, with the solid and the dashed curves indicating the 1-$\sigma$ and 2-$\sigma$ contours. The true oblateness and obliquity of Jupiter and Saturn are also shown for reference.}
    \label{fig:corner}
\end{figure*}

Our modeling procedure follows closely that of \citet{Liu:2024}. 
%The transit light curve is fit by the oblate transit model with the quadratic limb darkening law. The spherical transit model uses the \citet{Mandel:2002} 
We use \texttt{JoJo} to compute the transit light curve caused by an oblate planet, in which a quadratic limb darkening profile has been assumed. To describe the single transit event, the modeling parameters include the mid-transit time $t_0$, the impact parameter $b$, the effective planet-to-star radius ratio $R_{\rm mean}/R_\star$, the stellar density parameter $\rho_\star$ \citep[corresponding to the stellar mean density for a circular orbit,][]{Seager:2003}, the projected oblateness $f_\perp$, the projected obliquity $\theta_\perp$, and the two quadratic limb-darkening parameters $q_1$ and $q_2$ \citep{Kipping:2013}. The orbital period has been fixed at $P=130.1845$ days, which is the best-fit parameter of \cite{Libby-Roberts:2020}. For illustration purposes, we have also fit the data to the standard model \citep{Mandel:2002}.

The JWST light curve shows signatures of spot crossings, with the most significant one near the mid-transit. Instead of modeling the spot crossing event, we exclude a total duration of 3.12 hr starting from BJD=2460121.803 in the light curve modeling. This spot crossing event and the young nature of the host star both indicate that the star is fairly active and spotty, as is also seen in the Kepler data. To account for the dilution effect due to the unocculted spots on the stellar surface, we include a dilution factor $\delta$ in the light curve modeling and impose a flat prior up to $1\%$. This is a fairly generous range, as the rotational modulation of Kepler-51 due to starspots is $\sim1\%$ in the optical \citep{McQuillan:2014} and expected to be smaller in the IR.
%the optical light curve observed spot-crossing event only has an amplitude of $\lesssim 0.2\%$.
We model the light curve baseline and the potential red noises with a Gaussian process (GP) via \texttt{celerite2} \citep{Foreman:2018}. The Matern-3/2 kernel is used in our baseline model, which evaluates the non-diagonal elements of the covariance matrix as
\begin{equation}
    K_{ij}=\sigma^2\left(1+\frac{\sqrt{3}\tau}{\lambda}\right)e^{-\sqrt{3}\tau/\lambda}, \quad (i \neq j).
\end{equation}
Here $\tau=|t_i-t_j|$, $\sigma$ and $\lambda$ describe the amplitude and typical time scale of GP, respectively. We use a jitter term to account for the white noise, so the diagonal elements of the covariance matrix are $K_{ii}=\sigma^2 + {\rm jitter}^2$.

Following \citet{Liu:2024}, we sample the posterior distributions via \texttt{dynesty}, which uses the nested sample method \citep{Speagle:2020}, and we sample $f_\perp \cos(2\theta_\perp)$ and $f_\perp \sin(2\theta_\perp)$ rather than $f_\perp$ and $\theta_\perp$ directly. We consider this re-parameterization necessary because $\theta_\perp$ is ill-defined when $f_\perp=0$ and the posterior distribution would be highly non-Gaussian if one is to sample the $f_\perp$ vs.\ $\theta_\perp$ space directly \citep[e.g.,][]{Zhu:2014, Lammers:2024}. The periodic behaviour of $\theta_\perp$ is also circumvented with this re-parameterization, as $\theta_\perp$ is defined within $-90^\circ$ and $90^\circ$. The priors on $f_\perp \cos(2\theta_\perp)$ and $f_\perp \sin(2\theta_\perp)$ are set to be flat distributions between (-0.25, 0.25). To impose a flat prior on the oblateness parameter $f_\perp$, we then include a rectification term, $-\ln{f_\perp}$, into the total likelihood evaluation. For the other parameters, we adopt flat priors with rather generous ranges. For instance, the priors on the limb darkening parameters are flat distributions between zero and one.

As will be shown in Section~\ref{sec:result}, the oblateness parameters are strongly correlated with the stellar density parameter $\rho_\star$, which measures the transit duration. For eccentric orbit, $\rho_\star$ is related to the bulk density of the host star, $\bar{\rho}_\star$, the orbital eccentricity, $e$, and the argument of periapsis, $\omega$, through \citep{Dawson:2012}
\begin{equation} \label{eqn:rho_star}
    \rho_\star = \bar{\rho}_\star \left( \frac{1+e\sin\omega}{\sqrt{1-e^2}} \right)^{-3} .
\end{equation}
For Kepler-51d, the orbital eccentricity is constrained through transit timing variation modeling to be around 1\% (\citealt{Libby-Roberts:2020}; see also \citealt{Masuda:2014}). This would lead to a deviation of $<3\%$ from the stellar mean density $\bar{\rho}_\star$, which is almost negligible compared to the fractional uncertainty in $\bar{\rho}_\star$ alone. Therefore, we ignore the photo-eccentric factor and adopt as the Gaussian prior on $\rho_\star$ the measured value of $\lg[\bar{\rho}_\star/({\rm g\,cm^{-3}})]=0.307 \pm 0.017$ \citep{Libby-Roberts:2020}.

The above procedure defines our baseline model. As is detailed in Section~\ref{sec:result}, we have also explored several variations in the modeling procedure in order to confirm that our results are robust. These include different treatment of the $\rho_\star$ prior, different detrending methods, different GP kernels, the exclusion of the dilution factor, and modeling two-channel light curves rather than the single white light curve.

%%%%%%%%%%%%%%%%%%%%%%%%%%%
\section{Result} \label{sec:result}

\begin{figure*}
    \centering
    \includegraphics[width=0.45\linewidth]{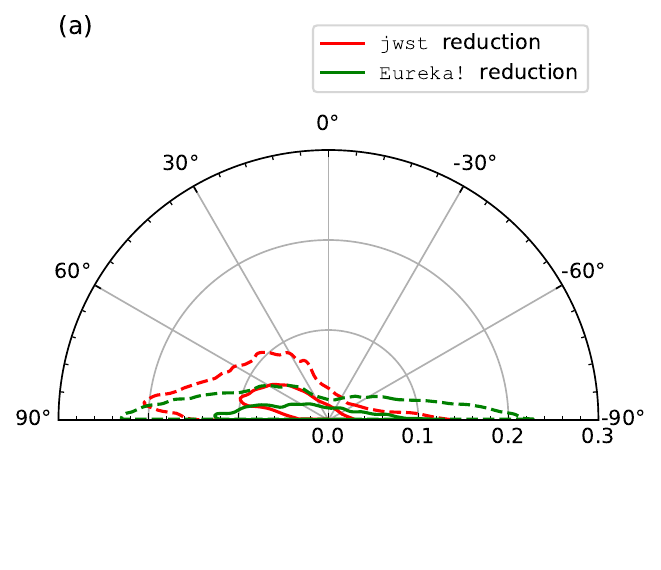}
    \includegraphics[width=0.45\linewidth]{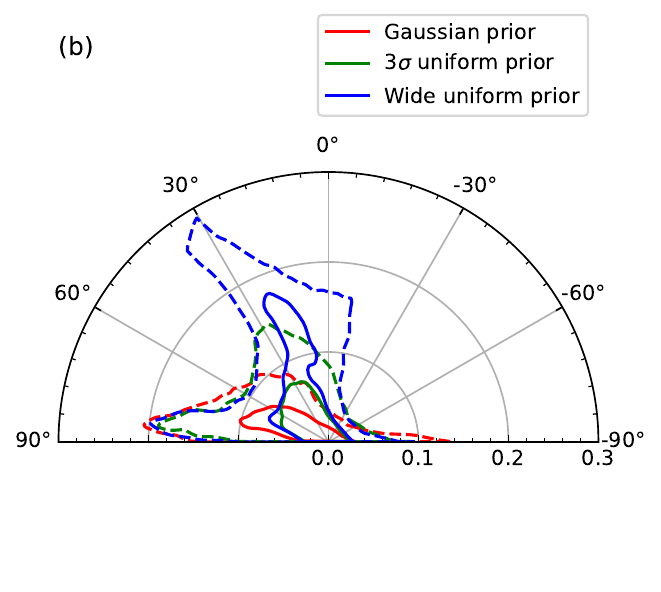}
    \includegraphics[width=0.45\linewidth]{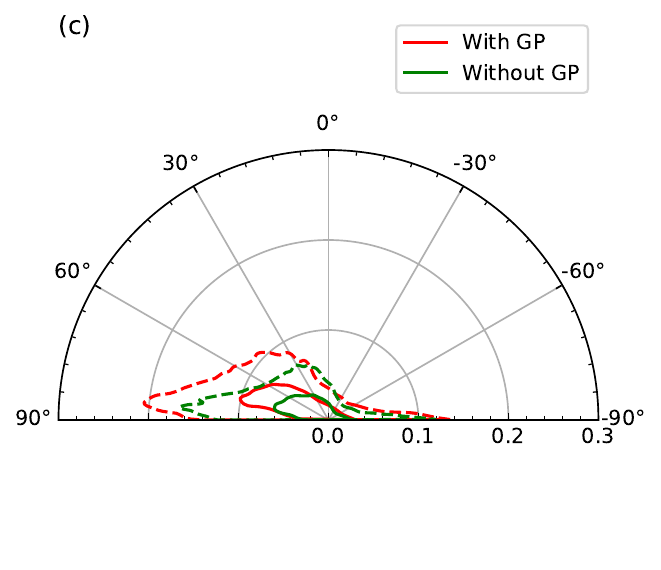}
    \includegraphics[width=0.45\linewidth]{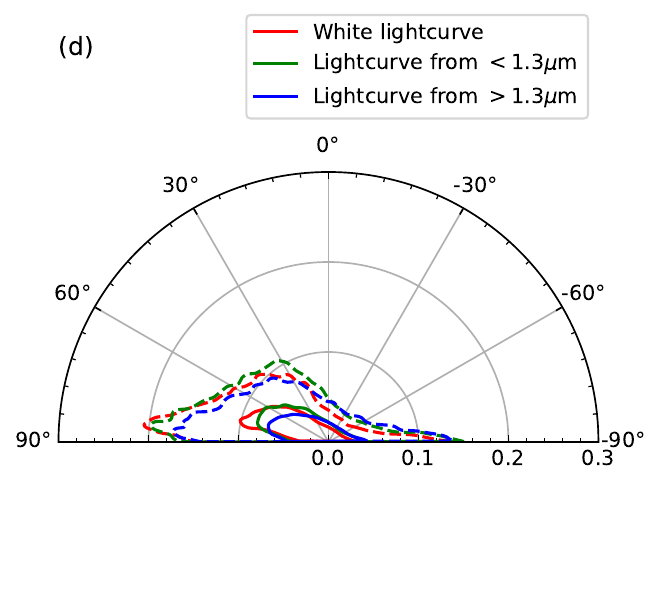}
    \caption{Constraints on the projected oblateness $f_\perp$ (along the radial direction) and the projected obliquity $\theta_\perp$ (along the azimuthal direction) in the polar coordinates. Different panels show the various consistency tests: panel (a) shows the results from two different reduction methods, panel (b) shows the results from imposing different $\lg\rho_\star$ priors in the modeling, panel (c) shows the results from using Gaussian process or not in the modeling, and panel (d) shows the results from modeling the two-channel light curves (lower panel of Figure~\ref{fig:all_lc}). The red contours are the results from our baseline model and thus the same in all panels. Similar to Figure~\ref{fig:corner}, the solid and dashed curves are the 1-$\sigma$ and 2-$\sigma$ contours, respectively.}
    \label{fig:tests}
\end{figure*}

We present in Figure~\ref{fig:lightcurve_baseline} the JWST transit light curve and the best-fit oblate and spherical models. The best-fit spherical model is shown only for the illustration purpose. As described in Section~\ref{sec:model}, a Gaussian process with the Matern-3/2 kernel has been included in this baseline model to model the systematic trend as well as the correlated noise. According to the middle panel of Figure~\ref{fig:lightcurve_baseline}, the systematic trend is at the level of $\sim 200\,$ppm, which is comparable to the level of scattering in the full model residuals (bottom panel of Figure~\ref{fig:lightcurve_baseline}). 

The corner plot of results from our baseline model is shown in Figure~\ref{fig:corner}. All parameters except for the dilution factor $\delta$ are reasonably constrained. Regarding $\delta$, although it is not constrained within the allowed range, it does not seem to correlate significantly with the other modeling parameters---in particular the oblateness parameters, namely $f_\perp \cos(2\theta_\perp)$ and $f_\perp \sin(2\theta_\perp)$. Therefore, we do not further investigate the cause and impact of the dilution factor for the purpose of the present work. The two oblateness parameters, $f_\perp \cos(2\theta_\perp)$ and $f_\perp \sin(2\theta_\perp)$, both show peak at zero, indicating that Kepler-51d is consistent with being perfectly spherical. In order to obtain an upper limit on the planetary oblateness, the posterior distributions of the two oblateness parameters then matter. Unlike the posterior distribution of $f_\perp \sin(2\theta_\perp)$, which is well constrained and has a small uncertainty $\sim 0.03$, the posterior distribution of $f_\perp \cos(2\theta_\perp)$ shows a long tail toward negative values, which yields a much weaker constraint on the projected oblateness $f_\perp$ along obliquities of $\pm 90^\circ$ (i.e., a Uranus-like obliquity). 

The 2D posterior distribution of the projected oblateness and obliquity recovered from the re-parameterized oblateness parameters is shown in the upper right corner of Figure~\ref{fig:corner}. The contours appear slightly tilted towards positive obliquity angles, but this may just be due to the statistical or systematic noises. For example, as shown in panel (a) of Figure~\ref{fig:tests}, if the white light curve from the \texttt{Eureka!} reduction is used, the 2D posterior distribution is slightly modified and appear more symmetric with respect to $\theta_\perp=0^\circ$. 
Regardless of the reduction method and obliquity angles, a projected oblateness up to $f_\perp\approx 0.2$ is ruled out at 2-$\sigma$ level. If limited to relatively small obliquity angles, such as $|\theta_\perp|\lesssim 60^\circ$, then we are able to put even more stringent upper limit on the projected oblateness $f_\perp$. In particular, for nearly zero obliquity angles ($\theta_\perp\lesssim 30^\circ$, which correspond to the spin--orbit aligned configurations, we are able to rule out $f_\perp \geq 0.1$ at the 2-$\sigma$ confidence level. In other words, the oblateness--obliquity combinations similar to our Saturn and Jupiter are highly disfavored.

As shown in Figure~\ref{fig:corner}, the long tail in the posterior distribution of the oblateness parameter $f_\perp \cos(2\theta_\perp)$ is primarily due to its correlation with the density parameter $\lg \rho_\star$, and thus the exact choice of the $\lg \rho_\star$ prior affects the resulting constraints on the planet oblateness (and obliquity). For example, if we adopt a prior that is uniform within the 3$\sigma$ range of the reported values of \citet{Libby-Roberts:2020}, then the derived joint posterior distribution of $f_\perp$ and $\theta_\perp$ is slightly modified. The results from this test are shown in panel (b) of Figure~\ref{fig:tests}. When a different $\lg\rho_\star$ prior is used, a higher upper limit on $f_\perp$ is obtained along the $\theta_\perp \approx 0^\circ$ direction, accompanied by a tighter upper limit on $f_\perp$ along the $\theta_\perp \approx \pm 90^\circ$ direction. Both constraints are significantly better than what one gets if imposes essentially no constraint on $\lg\rho_\star$ (i.e., a uniform prior between 0 and 1).
Therefore, the information of $\lg\rho_\star$ is crucial to determine $f_\perp$ when the spin obliquity is close to zero, in which case the oblateness signal is symmetric with respect to the mid-transit time and thus most difficult to recover/constrain (see also \citealt{Liu:2024}). This exercise suggests that, when one is to identify/select targets for oblateness constraints/detections with JWST, it should be taken into account if those targets have well-determined $\rho_\star$ parameters (including the photo-eccentric factor). We will return to this point in Section~\ref{sec:rho-effect}.

Our baseline model includes a Gaussian process with the Matern-3/2 kernel to model the systematic trend and the correlated noise. As shown in Figure~\ref{fig:corner}, the GP parameters are all well-constrained. Combining the GP amplitude $\sigma$ and the jitter term, we get the amplitude of the total point-to-point scattering of $663\,$ppm. This is comparable to the rough out-of-transit variation that is estimated in Section~\ref{sec:data} (see also Figure~\ref{fig:all_lc}). Concerning the correlated noises in the JWST light curve, the characteristic amplitude is constrained to be $\sigma \approx 100\,$ppm, which is substantially small compared to the total amplitude of the white noise term. This correlation also decreases at the timescale of $\tau \approx 1\,$hr, which is slightly shorter than the duration of the transit ingress/egress. Indeed, when we ignore the correlated noises by detrending the data with a quadratic polynomial and then fitting an oblate model to the normalized light curve, we find very similar constraints on the oblateness parameters, as shown in the panel (c) of Figure \ref{fig:tests}. Therefore, we conclude that the correlated noises are negligible for the purpose of constraining oblateness.

The final test we have performed is the oblateness modeling on the two-channel light curves, separated at the wavelength of $\sim1.3\,\mu$m (see Section~\ref{sec:data}). In this test, we conduct the oblateness modeling with the same settings as in our baseline model to the shorter-wavelength and longer-wavelength light curves separately, and the results are shown in the panel (d) of Figure~\ref{fig:tests}. The 2D posteriors from the two-channel light curves overlap largely with the one from the baseline model. This indicates that the oblateness constraint is not wavelength-dependent, which is consistent with our expectation. It is noteworthy that the 2D posterior distribution from either of the two-channel light curves is very similar to the one from the full white light curve, even though the two-channel light curves are individually noisier. This suggests that a full model of the spectroscopic light curve may be able to improve on the oblateness constraint, which we defer to a future study.

%%%%%%%%%%%%%%%%%%%%%%%%%%%
\section{Discussion} \label{sec:discussion}

\subsection{Spin period of the super-puff Kepler-51d}

Adopting an intermediate value for the dimensionless moment of inertia $\bar{C}=0.3$, we reduce the Darwin--Radau relation (Equation~\ref{eqn:darwin}) to
\begin{equation} %\label{eqn:darwin-radau}
    \frac{\Omega}{\Omega_{\rm brk}} \approx 0.34 \left(\frac{f}{0.1}\right)^{1/2} .
\end{equation}
If the two rather extreme values 0.2 and 0.4 are adopted, the prefactor changes fairly modestly from 0.34 to 0.40 and 0.28, respectively.

The projected oblateness that is constrained from the transit light curve serves as a lower limit on the true oblateness. Although the true oblateness can be statistically inferred by assuming an isotropic distribution for the spin obliquity \citep[e.g.,][]{Lammers:2024}, it is not directly constrained by the data. We choose to take the constraint on the projected oblateness as the constraint on the true oblateness. Adopting $f_\perp \leq 0.08$, which corresponds to the 2-$\sigma$ upper limit for $|\theta_\perp|\leq 30^\circ$ according to our baseline model, we are able to constrain the dimensionless spin rate $\Omega/\Omega_{\rm brk} \leq 0.3$. In other words, Kepler-51d, if in a nearly spin--orbit aligned configuration, is spinning no faster than $\sim30\%$ of its break-up spin rate. For references, Jupiter and Saturn spin at $28\%$ and $36\%$ of their break-up spin rates, respectively.
If the spin of Kepler-51d is substantially misaligned ($\gtrsim 30^\circ$) with respect to its orbital direction, then the planet may have an oblateness up to $\sim0.2$ and spin at $\lesssim 50\%$ of the break-up spin rate.

Kepler-51d has both mass and radius measurements, so we are able to set limits on its absolute rotation period. The break-up rotation period is given by
\begin{equation}
    P_{\rm brk} \equiv \frac{2\pi}{\Omega_{\rm brk}} \approx 16\,{\rm hr} \left(\frac{M_{\rm p}}{6\,M_\oplus}\right)^{-1/2} \left(\frac{R_{\rm p}}{9\,R_\oplus}\right)^{3/2} ,
\end{equation}
where we have adopted the updated mass and radius measurements of Kepler-51d from \citet{Masuda:2024}. Combining with the constraints on the dimensionless spin rate, we conclude that the rotation period of Kepler-51d is $\gtrsim53\,$hr if its spin aligns with the orbital direction to within $30^\circ$, or $\gtrsim33\,$hr if its spin direction is highly misaligned. Therefore, the rotation period of Kepler-51d is longer than the rotation period of any of the four giant planets in our Solar System.

With $M_{\rm p}\approx 6\,M_\oplus$ and $a\approx 0.5\,$au \citep{Masuda:2014}, Kepler-51d is not expected to have been tidally spun down \citep[e.g.,][]{Murray:1999}. Nor is it massive enough to have opened a gap in the protoplanetary disk \citep{Lin:1986}, which makes it different from the formation of Jupiter and Saturn. The near resonant configuration of the Kepler-51 system also suggests that the planets most likely did not experience giant impacts after their formation, which, if occurred, might have significantly modified their spin rates and orientations \citep[e.g.,][]{LiLai:2020}, so it is also different from Uranus and Neptune \citep{Ida:2020}.
Therefore, the accretion of mass and angular momentum onto Kepler-51d during its formation may have determined its rotation state, and, given the known mass and orbital separation of Kepler-51d, this accretion is three-dimensional and presumably in the relative cold region \citep{Lee:2016}.

No theoretical work has studied the spin state of super-puff planets, so we turn to the works that looked into the spin evolution of more typical low-mass planets. Based on three-dimensional hydrodynamical simulations, \citet{Takaoka:2023} studied the spins of relatively low-mass protoplanets that grew via pebble accretion, the leading theory explaining the formation of typical Kepler planets including super-puffs \citep{Ormel:2010, Lee:2016}. That study shows that a protoplanet should spin in prograde direction at nearly the break-up rotation if its mass approaches the rotation-induced isolation mass, given by $M_{\rm iso, rot}\approx 1.2 (a/{\rm au})^{1/4}\,M_\oplus$. The mass of Kepler-51d almost certainly exceeds $M_{\rm iso, rot}$, unless the planetary mass has been substantially overestimated and/or the planet was born beyond $\gtrsim200\,$au. The growth beyond this isolation mass is not discussed in \citet{Takaoka:2023}, but unless some unknown physical processes can efficiently reduce the angular momentum of the protoplanet, a fast spin would still be expected for a planet with mass above the aforementioned isolation mass. Our constraint that the spin rate of Kepler-51d is $\lesssim 30\%$ (for nearly spin-orbit aligned configurations) may therefore be in tension with the theoretical prediction, although more works are certainly needed in order to link the theoretical models and the observational constraints.

\subsection{Implications to planet oblateness detections with JWST} \label{sec:rho-effect}

The degeneracy between $\lg\rho_\star$, which effectively measures the transit duration, and the oblateness parameters is the bottleneck that limits our better constraining the planet oblateness. This degeneracy is especially prominent when the planetary spin obliquity is close to zero. While planetary spins can be tilted relative to their orbital direction (see \citealt{Liu:2024} and references therein), it remains the most natural outcome of the giant planet formation theories that the planetary spin is aligned with its orbital direction. Therefore, in order to constrain or even directly detect the planet oblateness with the precious JWST observations, it is important to select targets whose stellar density parameter (Equation~\ref{eqn:rho_star}) is known to a good precision.

What precision of $\rho_\star$ is required then? At zero obliquity, an oblate planet produces a longer transit duration by the fraction
\begin{equation}
    \frac{\delta \tau}{\tau} \approx \frac{(R_{\rm eq}-R_{\rm mean})}{\sqrt{1-b^2} R_\star} \approx \frac{f}{2\sqrt{1-b^2}} \frac{R_{\rm mean}}{R_\star} .
\end{equation}
In the latter step we have made use of $R_{\rm mean}=\sqrt{1-f}R_{\rm eq}$ and $f\ll 1$. Evaluating at the typical values ($f=0.1$ and $R_{\rm mean}/R_\star=0.1$) and ignoring the impact parameter factor, we get $\delta \tau/\tau \approx 0.5\%$. Furthermore, given that $\tau \propto \rho_\star^{-1/3}$, to constrain the transit duration to a precision of $\sim0.5\%$ seems to require to know the density parameter $\rho_\star$ to within $\sim1.5\%$. However, in the above estimation we have only considered the variation in the transit duration due to an oblate planet. A spin--orbit aligned oblate planet will also modify the slopes of the ingress and egress light curves, which, although has the price of introducing the degeneracy with the impact parameter $b$, can also be used to constrain the oblateness. Considering that the $\sim5\%$ precision of the density parameter $\rho_\star$ in the case of Kepler-51d is able to improve the oblateness constraint substantially, we estimate that a $\lesssim 10\%$ precision in $\rho_\star$ remains useful.

The density parameter $\rho_\star$ involves not only the stellar mean density $\bar{\rho}_\star$ but also the orbital properties of the planet (namely $e$ and $\omega$; see Equation~\ref{eqn:rho_star}). With the precise photometric, spectroscopic, and astrometric measurements, the stellar mean density can usually be constrained at a few percent level (comparable to the systematic uncertainty, \citealt{Tayar:2022}), but the orbital parameters are typically not, especially for the long-period transiting giants that are suitable for oblateness detections with JWST. For example, Kepler-167e is one of the best targets based on its long orbital period and deep transit, as illustrated in the Figure~2 of \citep{Liu:2024}, but its orbital eccentricity is only constrained to $<0.29$ \citep{Chachan:2022}, which leads to a much larger uncertainty in the density parameter and consequently in the oblateness at near zero obliquity angles.

Our analysis therefore highlights the importance of identifying well-characterized planets for oblateness measurements with JWST, as well as the importance to better characterize the known giant planets at wide orbits. 

\subsection{Comparison with \citet{Lammers:2024}} \label{sec:comparison}

Our results presented here are in general good agreement with the work by \citet{Lammers:2024}, which also analyzed the same JWST transit data of Kepler-51d in the context of oblateness constraint. The two independent analyses both find that Kepler-51d does not appear oblate in its projected shape, which implies that this extremely low-density planet probably spins relatively slowly.

Our work differs from that of \citet{Lammers:2024} in several aspects. The most notable difference is that we have taken into account the dilution effect due to starspots and performed multiple tests, as detailed in Section~\ref{sec:result}, in order to confirm the robustness of the results.In addition, our reduced white light curve covers a broader wavelength range and has smaller noises than that of \citet{Lammers:2024}. The latter used the wavelength range of 0.68--2.3\,$\mu$m and had an out-of-transit variation of 870\,ppm.The modeling process also differs in several ways, including the light curve computing algorithm, the modeling parameters, the posterior sampling method, and the use of GP to model correlated noises.

\citet{Lammers:2024} also investigated the possibility that the extremely low density of Kepler-51d could be produced by a more normal planet with a nearly face-on ring \citep{Piro:2020}. This scenario is not considered in the present work.

\section{Summary} \label{sec:summary}

%We summarize our results in a few sentences here...
In this work, we present constraints on the projected shape of the extremely low-density planet, Kepler-51d, using the transit light curve that was observed by JWST.
We show that the projected shape of Kepler-51d is consistent with being spherical and exclude an oblateness $f_\perp>0.08$, if the planetary spin is aligned with its orbital direction to within $30^\circ$, and $f_\perp>0.2$, if the planetary spin is even more misaligned. 
We have conducted multiple tests to show that the above constraints are robust against the different data reduction methods or modeling procedures.
Taking these as the limit on the true shape of the planet, we constrain the rotation rate to $\lesssim30\%$ ($\lesssim 50\%$) of its break-up spin rate and, given the measured mass and radius, a rotation period $\gtrsim 53\,$hr ($\gtrsim 33\,$hr) for the nearly aligned (largely misaligned) spin configuration. This extremely low-density planet is therefore spinning slowly. 
The spin of Kepler-51d is relatively slow, if compared to the extrapolation from the theoretical models that estimated the spin rate of low-mass planets. 
Our work also highlights the importance of better characterizations of the host star and planetary orbit in order to better constrain and eventually detect the oblateness of transiting exoplanets.

\begin{acknowledgments}
We would like to thank Fei Dai, Zhecheng Hu, Yu Wang, and Yifan Zhou for discussions. 
We also thank the anonymous referee for comments and suggestions on the manuscript.
This work is supported by the National Natural Science Foundation of China (grant No.\ 12173021 and 12133005). 
This work is based on observations made with the NASA/ESA/CSA James Webb Space Telescope. The JWST data presented in this article were obtained from the Mikulski Archive for Space Telescopes (MAST) at the Space Telescope Science Institute, which is operated by the Association of Universities for Research in Astronomy, Inc., under NASA contract NAS 5-03127 for JWST. These observations are associated with the program \#2571. The specific observations analyzed can be accessed via \dataset[doi: 10.17909/czps-pt89]{https://doi.org/10.17909/czps-pt89}.
The Center for Exoplanets and Habitable Worlds is supported by the Pennsylvania State University, the Eberly College of Science, and the Pennsylvania Space Grant Consortium. Part of the computations for this research were performed on the Pennsylvania State University’s Institute for Computational and Data Sciences Advanced CyberInfrastructure (ICDS-ACI).  This content is solely the responsibility of the authors and does not necessarily represent the views of the Institute for Computational and Data Sciences.
CIC acknowledges support by NASA Headquarters through an appointment to the NASA Postdoctoral Program at the Goddard Space Flight Center, administered by ORAU through a contract with NASA.

\end{acknowledgments}

\bibliography{51d}{}
\bibliographystyle{aasjournal}

\end{document}